\documentclass[preprint]{elsarticle}

\usepackage{lineno,hyperref}
\modulolinenumbers[5]

\journal{Journal of \LaTeX\ Templates}

\usepackage{graphicx}
\usepackage{amsmath}
\usepackage{amsthm}
\usepackage{appendix}
\usepackage{algorithm}
\usepackage{algpseudocode}
\usepackage{booktabs}

\theoremstyle{definition}

\begin{document}

\begin{frontmatter}

\title{Geodesic curves in Gaussian random field manifolds}

\author{Alexandre L. M. Levada}
\address{Computing Department, Federal Universisty of S\~ao Carlos, SP, Brazil}
\ead{alexandre.levada@ufscar.br}

\begin{abstract}
Random fields are mathematical structures used to model the spatial interaction of random variables along time, with applications ranging  from statistical physics and thermodynamics to system's biology and the simulation of complex systems. Despite being studied since the 19th century, little is known about how the dynamics of random fields are related to the geometric properties of their parametric spaces. For example, how can we quantify the similarity between two random fields operating in different regimes using an intrinsic measure? In this paper, we propose a numerical method for the computation of geodesic distances in Gaussian random field manifolds. First, we derive the metric tensor of the underlying parametric space (the $3 \times 3$ first-order Fisher information matrix), then we derive the 27 Christoffel symbols required in the definition of the system of non-linear differential equations whose solution is a geodesic curve starting at the initial conditions. The fourth-order Runge-Kutta method is applied to numerically solve the non-linear system through an iterative approach. The obtained results show that the proposed method can estimate the geodesic distances for several different initial conditions. Besides, the results reveal an interesting pattern: in several cases, the geodesic curve obtained by reversing the system of differential equations in time does not match the original curve, suggesting the existence of irreversible geometric deformations in the trajectory of a moving reference traveling along a geodesic curve. 
\end{abstract}


\end{frontmatter}


\section{Introduction}

In physics and applied mathematics, a fundamental question in the study of complex systems is: how can we measure an intrinsic distance between two systems operating in different regimes? In complex networks, different approaches have been proposed, from graph Laplacian based distances \cite{Graph_dist} to persistent entropy between homological groups \cite{Frontiers}. However, little is known about intrinsic distance functions between stochastic systems modeled by random fields \cite{Entropy}. Often, researchers extract several features from these systems and adopt an extrinsic similarity measure, such as the Euclidean distance, the Minkowski distance or the cosine similarity \cite{similarity}. However, it has been observed that the parametric space of most physical systems define a manifold, which, roughly speaking, is a curved space which locally resembles a ``flat'' Euclidean space \cite{Physical_Systems,Tristan}. Therefore, the use of extrinsic distance functions is not suitable since usually the shortest path (straight line) does not belong to the underlying manifold itself.

Information geometry is a recent research field that combine efforts in differential geometry and information theory to study the intrinsic geometry of the parametric spaces of random variables \cite{Amari2021}. It has been shown that the parametric spaces of independent random variables whose probability density functions belong to the exponential family of distributions exhibit constant curvature \cite{Dodson}. One example is the Gaussian density, in which the parametric space has constant negative curvature (hyperbolic geometry) \cite{Pinele}. Several works in literature report on geodesic distances between independent and identically distributed random variables from the exponential family. But when a spatial dependence structure emerges from the local interaction of these random variables, the scenario is not completely known. 

In this paper, our focus are Gaussian random fields, which are mathematical models employed in the study of non-deterministic phenomena in which non-linear interactions between random variables lead to the emergence of long range correlations, complexity and phase transitions \cite{GaussianRandomFields}. In opposition to other random field models such as the Ising and q-state Potts model, Gaussian random fields are composed by variables that can assume an infinite number of states, that is, the set of possible states is continuous. In order to improve the mathematical tractability and to simplify the derivations of the metric tensor and the Chistoffel symbols of the metric \cite{ChristoffelSymbols}, we assume some some simplifying hypothesis. First, we only model binary relationships, which means that we have a pairwise interaction model. And second, we consider that the inverse temperature parameter, which is responsible for controlling the spatial dependence structure along the field, is spatially invariant and isotropic.

The main contributions of this paper are: 1) to derive the exact metric tensor of pairwise isotropic Gaussian-Markov random field manifolds; 2) to derive the Christoffel symbols of the metric; 3) to derive the system of nonlinear differential equations that govern the dynamics of a moving reference along geodesic curves in the underlying manifold; 4) to propose a computational method to compute the geodesic distance between two pairwise isotropic Gaussian-Markov random fields using the the Runge-Kutta \cite{RK4} method and Markov-Chain Monte Carlo simulation \cite{MCMC2021}; and 5) to compare the Euclidean and geodesic distances between different pairwise isotropic Gaussian-Markov random field models. Computational experiments indicate that the proposed method provides a feasible approach for the numerical computation of geodesic curves in the underlying parametric space of Gaussian random fields.

The remaining of the paper is organized as follows: in Section 2, we present the isotropic pairwise Gaussian-Markov random field (GMRF) model and its pseudo-likelihood function. In Section 3, we present the mathematical derivation of the metric tensor of pairwise isotropic GMRF's manifolds. Section 4 presents the derivation of the Christoffel symbols of the metric, used to build the system of differential equations that define the geodesic curves along the manifold. Finally, in Section 5 we present our conclusions and final remarks.

\section{The random field model}

\begin{figure}[ht]
	\begin{center}
	\includegraphics[scale=0.4]{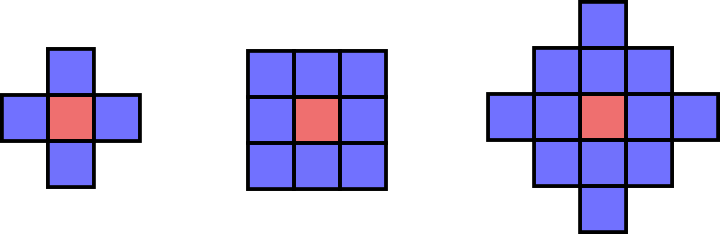}
	\end{center}
	\caption{First, second and third order neighborhood systems on a 2D lattice.}
	\label{fig:neigh}
\end{figure}

In our investigations, we adopt random fields composed by Gaussian random variables, which implies that the set of possible states at a given position is continuous \cite{GMRFBook}. In practice, each variable in the field can assume an infinite number of states. In order to make the mathematical derivations feasible, we assume some simplifying hypothesis: 1) the maximum clique size is two, that is, there are only binary relationships, which leads to a pairwise interaction random field; 2) the inverse temperature parameter, which controls the spatial dependence structure, is invariant along the random field and also isotropic, in the sense that the potentials for the neighboring interactions are the same in all directions of the field. Moreover, by invoking the Hammersley-Clifford theorem \cite{Hammersley}, which states the equivalence between Gibbs random fields (global models) and Markov random fields (local models), it is possible to characterize a the model as a pairwise isotropic Gaussian-Markov random field in terms of a set of local conditional density functions:

\begin{equation}
	p\left( x_{i} | \eta_{i}, \vec{\theta} \right) = \frac{1}{\sqrt{2\pi\sigma^2}}exp\left\{-\frac{1}{2\sigma^{2}} \left[ x_{i} - \mu - \beta \sum_{j \in \eta_{i}} \left( x_{j} - \mu \right) \right]^{2} \right\}
	\label{eq:GMRF}
\end{equation} where $\eta_i$ denotes the the second-order neighborhood system, comprised by the 8 nearest neighbors of $x_i$, $\vec{\theta} = (\mu, \sigma^{2}, \beta)$ denotes the vector of model parameters, with $\mu$ and $\sigma^{2}$ being, the expected value (mean) and the variance of the random variables in the lattice, and with $\beta$ being the inverse temperature. It is straightforward to see that when $\beta = 0$, the model degenerates to a regular Gaussian distribution where the variables become independent. The main advantage of using the local model is that we avoid the joint Gibbs distribution. Figure \ref{fig:neigh} shows the first, second and third order neighborhood systems defined on a 2D lattice.

Isotropic pairwise Gaussian-Markov random fields (GMRF's) are mathematical structures particularly suitable to study spatially dependent continuous random variables by means of non-linear interactions between local neighborhoods that are part of a lattice. The main advantage of this model in comparison to other random field models is related to mathematical tractability. Often, we cannot derive closed-form expressions for several information geometry measures, since they require the calculation of expectations. In pairwise isotropic GMRF's it is possible to compute these quantities without numerical approximations, which drastically reduces the computational burden of numerical simulations. An example is the exact closed-form expressions for the components of the metric tensor of the parametric space (Fisher information matrix) and the Christoffel symbols.

\subsection{The pseudo-likelihood function of Gaussian random fields}

In the mathematical statistics literature, it has been shown that the likelihood function of a model $p( \mathbf{X}| \vec{\theta} )$ can be expressed in terms of its natural parameters:

\begin{equation}
	p(\mathbf{X}| \vec{\theta} ) = exp\left\{\sum_{j=1}^{K}c_{j}( \vec{\theta} )T_{j}\left( \mathbf{X} \right) + d( \vec{\theta} ) + S\left( \mathbf{X} \right) \right\}
\end{equation} where $S\left(\mathbf{X} \right)$ is a function of the observations, $d(\vec{\theta})$ is a function of the parameters, $\vec{T} = \left( T_{1}\left(\mathbf{X} \right), T_{2}\left(\mathbf{X} \right), \ldots, T_{k}\left(\mathbf{X} \right) \right)$ denotes the vector of sufficient statistics and $\vec{c} = (c_{1}(\vec{\theta}), c_{2}(\vec{\theta}), \ldots, c_{k}(\vec{\theta}))$ denotes the vector of natural parameters. In pairwise isotropic Gaussian-Markov random fields, the pseudo-likelihood function, which is based on the conditional independence principle \cite{Besag}, is defined as the product of the local conditional density functions. Let $\mathbf{X} = \{x_{1}, x_{2}, \ldots, x_{n}\}$ be a sample of a pairwise isotropic Gaussian-Markov random field model where $\Delta=8$ represents the number of neighbors. Note that, as the number of natural parameters is greater than the number of parameters, isotropic pairwise GMRF's are curved models:

\begin{align}
	& F\left(\mathbf{X}| \eta_{i}, \vec{\theta} \right) = \left( 2\pi\sigma^2\right)^{-n/2}exp\left\{ -\frac{1}{2\sigma^2}\sum_{i=1}^{n}\left[ \left(x_{i} - \mu \right) - \beta \sum_{j\in\eta_i}\left( x_{j} - \mu \right)  \right]^2  \right\}
\end{align}

\begin{align}
	\nonumber & {} = exp\left\{ -\frac{n}{2}\left[ log(2\pi\sigma^2) + \frac{\mu^2}{\sigma^2} \right] + \frac{\beta\Delta\mu^2 n}{\sigma^2}\left[ 1 - \frac{\beta\Delta}{2} \right] \right\} \times \\ \nonumber 
	& \quad exp\left\{ \left[ \frac{\mu}{\sigma^2}\left(1 - \beta\Delta\right) \right]\sum_{i=1}^{n}x_{i} -\frac{1}{2\sigma^2}\sum_{i=1}^{n}x_{i}^2 + \frac{\beta}{\sigma^2}\sum_{i=1}^{n}\sum_{j\in\eta_i}x_{i}x_{j} \right. \nonumber \\ & \hspace{2cm} \left. - \left[ \frac{\beta\mu}{\sigma^2}(1 - \beta\Delta)\right]\sum_{i=1}^{n}\sum_{j\in\eta_i}x_{j} - \frac{\beta}{2\sigma^2}\sum_{i=1}^{n}\sum_{j\in\eta_i}\sum_{k\in\eta_i}x_{j}x_{k}  \right\} \nonumber
\end{align}

Looking at the pseudo-likelihood function, it is straightforward to identify the following correspondence:

\begin{align}
	\vec{c} & = \left( \left[ \frac{\mu}{\sigma^2}\left(1 - \beta\Delta\right) \right], -\frac{1}{2\sigma^2}, \frac{\beta}{\sigma^2}, -\left[ \frac{\beta\mu}{\sigma^2}(1 - \beta\Delta)\right], - \frac{\beta}{2\sigma^2} \right) \\ \nonumber
	\vec{T} & = \left( \sum_{i=1}^{n}x_{i}, \sum_{i=1}^{n}x_{i}^2, \sum_{i=1}^{n}\sum_{j\in\eta_i}x_{i}x_{j}, \sum_{i=1}^{n}\sum_{j\in\eta_i}x_{j}, \sum_{i=1}^{n}\sum_{j\in\eta_i}\sum_{k\in\eta_i}x_{j}x_{k}  \right) 
\end{align} with $S(\mathbf{X}) = 0$ and 

\begin{equation}
	d(\vec{\theta}) = -\frac{n}{2}\left[ log(2\pi\sigma^2) + \frac{\mu^2}{\sigma^2} \right] + \frac{\beta\Delta\mu^2 n}{\sigma^2}\left[ 1 - \frac{\beta\Delta}{2} \right]
\end{equation}

Moreover, note that if $\beta = 0$, the pseudo-likelihood function is reduced to the regular likelihood function of a Gaussian random variable, in which the number of parameters and the number of natural parameters are the same:

\begin{align}
	F\left(\mathbf{X}| \vec{\theta} \right) & = exp\left\{ \frac{\mu}{\sigma^2}\sum_{i=1}^{n}x_{i} -\frac{1}{2\sigma^2}\sum_{i=1}^{n}x_{i}^2 - \frac{n}{2}\left[ log(2\pi\sigma^2) + \frac{\mu^2}{\sigma^2} \right] \right\}
\end{align} where $S(\mathbf{X}) = 0$ and:

\begin{align}
	\vec{c} = \left( \frac{\mu}{\sigma^2}, -\frac{1}{2\sigma^2} \right) \quad
	\vec{T} = \left( \sum_{i=1}^{n}x_{i}, \sum_{i=1}^{n}x_{i}^2 \right) \quad
	d(\vec{\theta}) = -\frac{n}{2}\left[ log(2\pi\sigma^2) + \frac{\mu^2}{\sigma^2} \right]
\end{align}

These equations indicate that when the inverse temperature parameter increases, the model becomes curved, as the number of natural parameters becomes greater than the number of parameters. For $\beta=0$, the underlying parametric space is a surface with constant negative curvature. However, for larger values of $\beta$, the geometry of parametric space suffers drastic changes, as the surface slowly becomes a 3D manifold and the system undergoes a phase transition.

\section{The metric tensor of Gaussian random field manifolds}

Differential geometry provide the basis for the study of intrinsic properties of differentiable manifolds, a mathematical structure that locally resembles an Euclidean space, but globally exhibit curvature \cite{ElementaryDG,Bar,Manfredo,Tristan}. One fundamental concept in differential geometry is an homeomorphism. In summary, a map is $f: X \rightarrow Y$ is called a homeomorphism if and only if $f$ is a bijection and both $f$ and its inverse $f^{-1}$ are continuous functions. A manifold $M$ is a topological space with the property that each point has a neighborhood that is homeomorphic to an open subset of the n-dimensional Euclidean space $R^n$. A Riemannian manifold is equipped with a metric that allows the measurement of distances and angles. Such structure is known as the metric tensor of the manifold. 

\begin{figure}[ht]
	\begin{center}
	\includegraphics[scale=0.3]{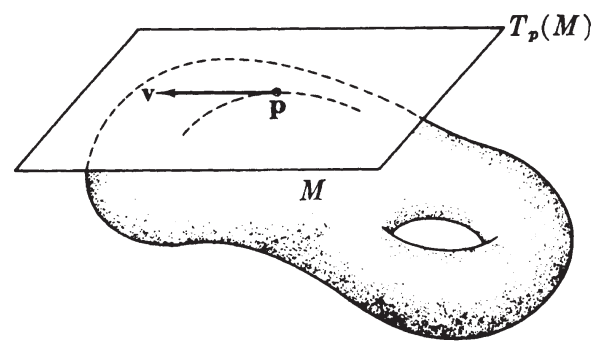}
	\end{center}
	\caption{A surface $M$ and the tangent plane $T_pM$.}
	\label{fig:tangent}
\end{figure}

It is possible to assign to each point $p$ in a manifold a linear vector space that is composed by all tangent vectors at $p$. This is a rather informal definition of the tangent space at $p$, which is an Euclidean space with the same dimension of the manifold. Figure \ref{fig:tangent}, reproduced from O'Neill's book \cite{ElementaryDG}, illustrates an arbitrary surface and the tangent plane at an arbitrary point $p \in M$.

Any living being on a manifold $M$ interested in studying its intrinsic geometry needs to know how to measure distance between two points and the angle between two vectors belonging to a tangent space. Note that this task is completely different from the distance between these points as measured by a living being that lives in the the ambient space, since, often, the extrinsic straight line between the points in $R^n$ will not be part of the manifold itself. The mathematical object that allows one to compute lengths on the manifold, and also angles and areas, is the metric tensor \cite{Pressley,Woodward}.

More precisely, a metric tensor at $p \in M$ is a function $g(\vec{x}_p, \vec{y}_p)$ which takes as inputs a pair of tangent vectors $\vec{x}_p$ and $\vec{y}_p$ and produces as an output a real number (scalar), with the following properties:

\begin{itemize}
	\item $g(\vec{x}_p, \vec{y}_p)$ is bilinear, that is, it is linear in each argument:
	\begin{align}
		g(a \vec{x}_p + b \vec{y}_p, \vec{z}_p) = a g(\vec{x}_p, \vec{z}_p) + b g(\vec{y}_p, \vec{z}_p) \\
		g(\vec{z}_p, a \vec{x}_p + b \vec{y}_p) = a g(\vec{z}_p, \vec{x}_p) + b g(\vec{z}_p, \vec{y}_p)
	\end{align}
	\item $g(\vec{x}_p, \vec{y}_p)$ is symmetric, that is:
	\begin{equation}
		g(\vec{x}_p, \vec{y}_p) = g(\vec{y}_p, \vec{x}_p)
	\end{equation}
	\item $g(\vec{x}_p, \vec{y}_p)$ is nondegenerate, that is, for every $\vec{x}_p \neq 0$, there exists a $\vec{y}_p$ such that $g(\vec{x}_p, \vec{y}_p) \neq 0$
\end{itemize}

The components of the metric tensor in any basis form the entries of a $k \times k$ symmetric matrix, where $k$ is the dimensionality of the tangent spaces.

\subsection{Information geometry}

It has been shown by information geometry that the Fisher information matrix is the metric tensor that equips the underlying parametric space of a statistical manifold \cite{Amari,Nielsen}. In practical terms, the metric tensor makes it possible to express the square of an infinitesimal displacement in the manifold, $ds^2$, as a function of an infinitesimal displacement in the tangent space, which in case of a simple 2D manifold is given by a vector $[du, dv]$. Assuming a matrix notation we have:

\begin{equation}
	ds^2 = \begin{bmatrix} du & dv \end{bmatrix} \begin{bmatrix} A & B \\ B & C \end{bmatrix}\begin{bmatrix} du \\ dv \end{bmatrix} = A du^2 + 2Bdudv + C dv^2
\end{equation} where the matrix of coefficients $A$, $B$, e $C$ is the metric tensor. If the metric tensor is a positive definite matrix, the manifold is is known as Riemannian. Note that in the Euclidean case, the metric tensor is the identity matrix (the space is flat), and we have the known Pythagorean relation $ds^2 = du^2 + dv^2$. 

\subsection{Fisher information}

Let $p(X;\vec{\theta})$ be a probability density function where $\vec{\theta} = (\theta_1, \ldots, \theta_n) \in \Theta$ is the vector of parameters. The Fisher information matrix, which is the natural Riemannian metric of the parametric space, is defined as:

\begin{equation}
	\left\{ I(\vec{\theta}) \right\}_{ij} = E\left[ \left(\frac{\partial}{\partial\theta_i} log~p(X; \vec{\theta}) \right)\left(\frac{\partial}{\partial\theta_j} log~p(X;\vec{\theta}) \right) \right], \text{~~~~ for } i,j=1,\ldots,n
\end{equation}

\subsection{Derivation of the metric tensor components}

In the following, we provide a complete discussion about the mathematical derivation of all components of the metric tensor of pairwise isotropic Gaussian-Markov random fields. As the parametric space is a 3D manifold, the Fisher information matrix has the shape:

\begin{equation}
	g(\vec{\theta}) = \left( \begin{array}{ccc}
	g_{11} & g_{12} & g_{13} \\ 
	g_{21} & g_{22} & g_{23} \\ 
	g_{31} & g_{32} & g_{33}
	\end{array} \right)
\end{equation}

For purposes of notation, we will denote that $\theta_1 = \mu$, $\theta_2 = \sigma^2$ and $\theta_3 = \beta$. Let us begin with the first component of the matrix, $g_{11}$, which involves the derivatives with respect to the $\mu$ parameter. Note that the first component of $g(\vec{\theta})$ is:

\begin{equation}
	g_{11} = I_{\mu\mu}(\vec{\theta}) = E\left[ \left(\frac{\partial}{\partial\mu} log~p(X; \vec{\theta}) \right)\left(\frac{\partial}{\partial\mu} log~p(X; \vec{\theta}) \right) \right]
\end{equation} where $p(X; \vec{\theta})$ is the replaced by the local conditional density function of the Gaussian random field, given by equation \eqref{eq:GMRF}. The computation of the derivatives leads to:

\begin{align}
	g_{11} & = E\left\{ \frac{1}{\sigma^2}\left(1 - \beta\Delta \right)^2 \frac{1}{\sigma^2}\left[ \left(x_i - \mu \right) - \beta \sum_{j\in\eta_i}\left(x_j - \mu \right) \right]^2 \right\} \label{eq:mu_mu_1} 
\end{align}

Expanding the square, we have:

\begin{align}
	g_{11} & = \frac{1}{\sigma^2}\left(1 - \beta\Delta \right)^2 E\left\{ \frac{1}{\sigma^2} \left[ \left(x_i - \mu\right)^2 - 2\beta\sum_{j\in\eta_i}\left( x_i - \mu \right)\left( x_j - \mu \right) \right. \right. \\ \nonumber & \hspace{4cm} \left. \left. + \beta^2 \sum_{j\in\eta_i}\sum_{k\in\eta_i}\left( x_j - \mu \right)\left(x_k - \mu  \right) \right] \right\} 
\end{align} 

And simplifying the expected values, we reach:

\begin{equation}
	g_{11} = \frac{\left(1 - \beta\Delta \right)^2}{\sigma^2} \left[ 1 - \frac{1}{\sigma^2}\left(  2\beta\sum_{j\in\eta_i}\sigma_{ij} - \beta^2\sum_{j\in\eta_i}\sum_{k\in\eta_i}\sigma_{jk} \right) \right]
\end{equation} where $\Delta$ is the cardinality of the neighborhood system ($\Delta = 8$ in a second-order system), $\sigma_{ij}$ is the covariance between the central variable $x_i$ and one of its neighbors $x_j \in \eta_i$ and $\sigma_{jk}$ is the covariance between two variables $x_j$ and $x_k$ belonging to the neighborhood $\eta_i$. The second component of the metric tensor is:

\begin{equation}
	g_{12} = I_{\mu\sigma^2}(\vec{\theta}) = E\left[ \left(\frac{\partial}{\partial\mu} log~p(X; \vec{\theta}) \right)\left(\frac{\partial}{\partial\sigma^2} log~p(X; \vec{\theta}) \right) \right]
\end{equation} which leads to:

\begin{align}
	\label{eq:g12}
	g_{12} & = \frac{(1 - \beta\Delta)}{2\sigma^6}E\left\{\left[ \left( x_i - \mu \right) - \beta\sum_{j\in\eta_i}\left(x_j - \mu \right) \right]^3 \right\} \\ \nonumber & \hspace{3cm} -\frac{(1 - \beta\Delta)}{2\sigma^4}E\left\{ \left( x_i - \mu \right) - \beta\sum_{j\in\eta_i}\left(x_j - \mu \right) \right\}	
\end{align}

Note that second term of equation \eqref{eq:g12} is zero, since:

\begin{equation}
	E\left[ x_i - \mu \right] - \beta\sum_{j\in\eta_i}E\left[ x_j - \mu\right] = 0 - 0 = 0
\end{equation} and the expansion of the first term of equation \eqref{eq:g12} leads to:

\begin{align}
	\label{eq:g12_exp}
	& E\left\{\left[ \left( x_i - \mu \right) - \beta\sum_{j\in\eta_i}\left(x_j - \mu \right) \right]^3 \right\} = E\left[ \left( x_i - \mu \right)^3 \right] \\ \nonumber & - 3\beta\sum_{j\in\eta_i}E\left[ (x_i - \mu) (x_i - \mu) (x_j - \mu) \right] \\ \nonumber & + 3\beta^2 \sum_{j\in\eta_i}\sum_{k\in\eta_i}E\left[ (x_i - \mu) (x_j - \mu) (x_k - \mu) \right] \\ \nonumber & - \beta^3 \sum_{j\in\eta_i}\sum_{k\in\eta_i}\sum_{l\in\eta_i}E\left( (x_j - \mu) (x_k - \mu) (x_l - \mu) \right]
\end{align}
	
The first term of \eqref{eq:g12_exp} is zero for Gaussian random variables, since every central moment of odd order is null. According to the Isserlis' theorem \cite{isserlis1918}, it is trivial to see that in fact all the third order cross terms are null, therefore, $g_{12} = 0$. The third component of the metric tensor is:

\begin{equation}
	g_{13} = I_{\mu\beta}(\vec{\theta}) = E\left[ \left(\frac{\partial}{\partial\mu} log~p(X; \vec{\theta}) \right)\left(\frac{\partial}{\partial\beta} log~p(X; \vec{\theta}) \right) \right]
\end{equation}

Plugging the local conditional density function and doing some basic algebra, we reach:

\begin{align}
	g_{13} = I_{\mu\beta}(\vec{\theta}) & = \frac{(1 - \beta\Delta)}{\sigma^4}\Bigg\{ E\left[ (x_i - \mu) (x_i - \mu) (x_j - \mu) \right]  \\ \nonumber & - 2\beta\sum_{j\in\eta_i}\sum_{k\in\eta_i}E\left[ (x_i - \mu) (x_j - \mu) (x_k - \mu) \right] \\ \nonumber & + \beta^2 \sum_{j\in\eta_i}\sum_{k\in\eta_i}\sum_{l\in\eta_i}E\left[ (x_j - \mu) (x_k - \mu) (x_l - \mu) \right] \Bigg\}
\end{align}

Once again, all the  third order moments are zero by the Isserlis's theorem, resulting in $g_{13}=0$. For the next component, by the symmetry of the metric tensor, $g_{21}=g{12}=0$. In order to calculate fifth component of the metric tensor, we have to compute:

\begin{equation}
	g_{22} = I_{\sigma^2 \sigma^2}(\vec{\theta}) = E\left[ \left(\frac{\partial}{\partial\sigma^2} log~p(X; \vec{\theta}) \right)\left(\frac{\partial}{\partial\sigma^2} log~p(X; \vec{\theta}) \right) \right]
\end{equation} which is given by:

\begin{align}
	g_{22} & = E\left\{ \left[ -\frac{1}{2\sigma^2} + \frac{1}{2\sigma^4}\left( x_i - \mu - \beta\sum_{j\in\eta_i}(x_j - \mu) \right) \right]^2 \right\} \\ \nonumber & = \frac{1}{4\sigma^4} - \frac{1}{2\sigma^6}E\left\{ \left[ (x_i - \mu) - \beta\sum_{j\in\eta_i}(x_j - \mu) \right]^2 \right\} \\ \nonumber & \hspace{1cm} + \frac{1}{4\sigma^8}E\left\{ \left[ (x_i - \mu) - \beta\sum_{j\in\eta_i}(x_j - \mu) \right]^4 \right\}
\end{align}

Note that the first expectation leads to the following equality:

\begin{equation}
	E\left\{ \left[ (x_i - \mu) - \beta\sum_{j\in\eta_i}(x_j - \mu) \right]^2 \right\} =  \sigma^2 - 2\beta\sum_{j\in\eta_i}\sigma_{ij} + \beta^2 \sum_{j\in\eta_i}\sum_{k\in\eta_i}\sigma_{jk} 
\end{equation}

For the second expectation, we have:

\begin{align}
	& E\left\{ \left[ (x_i - \mu) - \beta\sum_{j\in\eta_i}(x_j - \mu) \right]^4 \right\} = E\left[ (x_i - \mu)^4 \right] \\ \nonumber & \hspace{1cm} - 4\beta\sum_{j\in\eta_i}E\left[ (x_i - \mu)^3 (x_j - \mu) \right] \nonumber \\ & \hspace{1cm} + 6\beta^2 \sum_{j\in\eta_i}\sum_{k\in\eta_i}E\left[ (x_i - \mu)^2 (x_j - \mu) (x_k - \mu) \right] \\ \nonumber & \hspace{1cm} - 4\beta^3 \sum_{j\in\eta_i}\sum_{k\in\eta_i}\sum_{l\in\eta_i} E\left[(x_i - \mu) (x_j - \mu) (x_k - \mu) (x_l - \mu) \right] \\ \nonumber & \hspace{1cm} + \beta^4 \sum_{j\in\eta_i}\sum_{k\in\eta_i}\sum_{l\in\eta_i}\sum_{m\in\eta_i}E\left[(x_j - \mu) (x_k - \mu) (x_l - \mu) (x_m - \mu) \right]
\end{align} leading to five different expectation terms. We invoke the Isserlis' theorem for Gaussian random variables to express higher order moments in terms of second-order moments. Hence, after some algebraic manipulations, we have:

\begin{align}
	\label{eq:sigma_sigma_1}
	g_{22} & = \frac{1}{2\sigma^4} - \frac{1}{\sigma^6}\left[ 2\beta\sum_{j\in\eta_i}\sigma_{ij} - \beta^2 \sum_{j\in\eta_i}\sum_{k\in\eta_i}\sigma_{jk} \right] \\ \nonumber & + \frac{1}{\sigma^8}\left[ 3\beta^2 \sum_{j\in\eta_i}\sum_{k\in\eta_i}\sigma_{ij}\sigma_{ik} - \beta^3 \sum_{j\in\eta_i}\sum_{k\in\eta_i}\sum_{l\in\eta_i}\left( \sigma_{ij}\sigma_{kl} + \sigma_{ik}\sigma_{jl} + \sigma_{il}\sigma_{jk} \right) \right. \\ \nonumber & \hspace{1cm} \left. + \beta^4 \sum_{j\in\eta_i}\sum_{k\in\eta_i}\sum_{l\in\eta_i}\sum_{m\in\eta_i}\left( \sigma_{jk} \sigma_{lm} + \sigma_{jl}\sigma_{km} + \sigma_{jm}\sigma_{kl} \right)  \right] 
\end{align}

The sixth component of the metric tensor is given by:

\begin{equation}
	g_{23} = I_{\sigma^2 \beta}(\vec{\theta}) = E\left[ \left(\frac{\partial}{\partial\sigma^2} log~p(X; \vec{\theta}) \right)\left(\frac{\partial}{\partial\beta} log~p(X; \vec{\theta}) \right) \right]
\end{equation} which can be computed as:

\begin{align}
	 g_{23} = & E\left\{ \left[ -\frac{1}{2\sigma^2} + \frac{1}{2\sigma^4}\left( (x_i - \mu) - \beta\sum_{j\in\eta_i}(x_j - \mu)  \right)^2 \right] \times \right. \\ \nonumber & \hspace{2cm} \left. \left[ \frac{1}{\sigma^2}\left((x_i - \mu) - \beta\sum_{j\in\eta_i}(x_j - \mu)  \right)\left( \sum_{j\in\eta_i}(x_j - \mu) \right) \right] \right\}
\end{align}

\begin{align}	 
	\nonumber {} & = -\frac{1}{2\sigma^4} E\left\{  \left[ (x_i - \mu) - \beta\sum_{j\in\eta_i}(x_j - \mu) \right]\left[ \sum_{j\in\eta_i}(x_j - \mu) \right] \right\} \\ \nonumber & \hspace{2cm} + \frac{1}{2\sigma^6}E\left\{\left[ (x_i - \mu) - \beta\sum_{j\in\eta_i}(x_j - \mu) \right]^3 \left[ \sum_{j\in\eta_i}(x_j - \mu) \right]  \right\}
\end{align}

By computing the first expectation, we have:

\begin{equation}
	E\left\{  \left[ (x_i - \mu) - \beta\sum_{j\in\eta_i}(x_j - \mu) \right]\left[ \sum_{j\in\eta_i}(x_j - \mu) \right] \right\} = \sum_{j\in\eta_i}\sigma_{ij} - \beta\sum_{j\in\eta_i}\sum_{k\in\eta_i}\sigma_{jk}
\end{equation}

The expansion of the second expectation leads to:

\begin{align}
	& E\left\{\left[ (x_i - \mu) - \beta\sum_{j\in\eta_i}(x_j - \mu) \right]^3 \left[ \sum_{j\in\eta_i}(x_j - \mu) \right]  \right\} = \\ \nonumber & E\left\{ \left[ \sum_{j\in\eta_i}(x_j - \mu) \right] \left[ (x_i - \mu)^3 - 3\beta\sum_{j\in\eta_i}(x_i - \mu)^2 (x_j - \mu) \right. \right. \\ \nonumber \\ \nonumber & \hspace{4cm} \left. \left. + 3\beta^2 \sum_{j\in\eta_i}\sum_{k\in\eta_i}(x_i - \mu)(x_j - \mu)(x_k - \mu) \right. \right. \\ \nonumber  & \hspace{5cm} \left. \left. -\beta^3 \sum_{j\in\eta_i}\sum_{k\in\eta_i}\sum_{l\in\eta_i}(x_j - \mu)(x_k - \mu)(x_l - \mu) \right] \right\}
\end{align}

Again, by direct application of the Isserlis' theorem to express higher-order cross moments in terms of second-order moments and after some simplifications, we have:

\begin{align}
	\label{eq:sigma_beta_1}
	g_{23} & = \frac{1}{\sigma^4}\left[ \sum_{j\in\eta_i}\sigma_{ij} - \beta\sum_{j\in\eta_i}\sum_{k\in\eta_i}\sigma_{jk} \right] \\ \nonumber & - \frac{1}{2\sigma^6}\left[ 6\beta\sum_{j\in\eta_i}\sum_{k\in\eta_i}\sigma_{ij}\sigma_{ik} - 3 \beta^2 \sum_{j\in\eta_i}\sum_{k\in\eta_i}\sum_{l\in\eta_i}\left( \sigma_{ij}\sigma_{kl} + \sigma_{ik}\sigma_{jl} + \sigma_{il}\sigma_{jk} \right) \right. \\ \nonumber & \hspace{3cm} \left. + \beta^3 \sum_{j\in\eta_i}\sum_{k\in\eta_i}\sum_{l\in\eta_i}\sum_{m\in\eta_i} \left( \sigma_{jk}\sigma_{lm} + \sigma_{jl}\sigma_{km} + \sigma_{jm}\sigma_{kl} \right) \right]
\end{align}

It is straightforward to see that $g_{31}=g_{13}=0$ and $g_{32}=g_{23}$, since the metric tensor is symmetric. Finally, the last component is defined as:

\begin{equation}
	g_{33} = I_{\beta\beta}(\vec{\theta}) = E\left[ \left(\frac{\partial}{\partial\beta} log~p(X; \vec{\theta}) \right)\left(\frac{\partial}{\partial\beta} log~p(X; \vec{\theta}) \right) \right]
\end{equation} which is given by:

\begin{align}
	g_{33} & = \frac{1}{\sigma^4}E\left\{ \left[ (x_i - \mu) - \beta\sum_{j\in\eta_i}(x_j - \mu) \right]^2 \left[ \sum_{j\in\eta_i}(x_j - \mu) \right]^2  \right\} \\ \nonumber & = \frac{1}{\sigma^4} E\left\{ \left[ (x_i - \mu)^2 - 2\beta \sum_{j\in\eta_i} (x_i - \mu)(x_j - \mu) + \beta^2 \sum_{j\in\eta_i}\sum_{k\in\eta_i} (x_j - \mu)(x_k - \mu) \right] \times \right. \\ \nonumber & \left. \hspace{5cm} \left[ \sum_{j\in\eta_i}\sum_{k\in\eta_i} (x_j - \mu)(x_k - \mu) \right] \right\} \\ \nonumber & = \frac{1}{\sigma^4} E \left\{ \sum_{j\in\eta_i}\sum_{k\in\eta_i}(x_i - \mu)(x_i - \mu)(x_j - \mu)(x_k - \mu) \right. \\ \nonumber & \hspace{2cm} \left. - 2\beta\sum_{j\in\eta_i}\sum_{k\in\eta_i} \sum_{l\in\eta_i}(x_i - \mu)(x_j - \mu)(x_k - \mu)(x_l - \mu) \right. \\ \nonumber & \hspace{3cm} \left. + \beta^2 \sum_{j\in\eta_i} \sum_{k\in\eta_i} \sum_{l\in\eta_i} \sum_{m\in\eta_i}(x_j - \mu)(x_k - \mu) (x_l - \mu) (x_m - \mu)  \right\}
\end{align}
 
Once again, by using the Isserlis' formula and some algebra, we have:

\begin{align}
	\label{eq_beta_beta_1}
	g_{33} = \frac{1}{\sigma^2}\sum_{j\in\eta_i} \sum_{k\in\eta_i} \sigma_{jk} & + \frac{1}{\sigma^4} \left[ 2 \sum_{j\in\eta_i} \sum_{k\in\eta_i} \sigma_{ij} \sigma_{ik} \right. \\ \nonumber & \left. - 2\beta \sum_{j\in\eta_i} \sum_{k\in\eta_i} \sum_{l\in\eta_i} \left( \sigma_{ij}\sigma_{kl} + \sigma_{ik}\sigma_{jl} + \sigma_{il}\sigma_{jk} \right) \right. \\ \nonumber & \left. + \beta^2 \sum_{j\in\eta_i} \sum_{k\in\eta_i} \sum_{l\in\eta_i} \sum_{m\in\eta_i} \left( \sigma_{jk}\sigma_{lm} + \sigma_{jl}\sigma_{km} + \sigma_{jm}\sigma_{kl} \right) \right]
\end{align} concluding that the first fundamental form has the following structure:

\begin{equation}
	g(\vec{\theta}) = \left( \begin{array}{ccc}
	g_{11} & 0 & 0 \\ 
	0 & g_{22} & g_{23} \\ 
	0 & g_{32} & g_{33}
	\end{array} \right)
\end{equation} where $g_{23} = g_{32}$ and the non-zero elements are used to define how we compute an infinitesimal displacement in the manifold (parametric space) around the point $\vec{p} = (\mu, \sigma^2, \beta)$:

\begin{align}
	(ds)^2 & = \begin{bmatrix} d\mu & d\sigma^2 & d\beta \end{bmatrix}  \begin{bmatrix} g_{11} & 0 & 0 \\ 
	0 & g_{22} & g_{23} \\ 
	0 & g_{32} & g_{33} \end{bmatrix} \begin{bmatrix} d\mu \\ d\sigma^2 \\ d\beta \end{bmatrix} \\ \nonumber & = g_{11} (d\mu)^2 + g_{22} (d\sigma^2)^2 + g_{33} (d\beta)^2 + g_{23} (d\sigma^2)(d\beta) + g_{32} (d\beta)(d\sigma^2)
\end{align}

\subsection{The metric tensor components in tensorial notation}

In order to reduce the computational burden in the numerical computations, we propose to express the components of the first and second fundamental forms using Kronecker products (tensor products). First, note that we can convert each $3 \times 3$ neighborhood patch formed by $x_{i} \cup \eta_{i}$ into a vector $p_i$ of 9 elements by piling its rows. Then, we compute the covariance matrix of these vectors, for $i = 1, 2,..., n$ denoted by $\Sigma_{p}$. From this covariance matrix, we extract two main components: 1) a vector of size 8, $\vec{\rho}$, composed by the the elements of the central row of $\Sigma_{p}$, excluding the middle one, which denotes the variance of $x_i$ (we want only the covariances between $x_i$ and $x_j$, for $j \neq i$; and 2) a sub-matrix of dimensions $8 \times 8$, $\Sigma_{p}^{-}$, obtained by removing the central row and central column from $\Sigma_{p}$ (we want only the covariances between $x_j \in \eta_i$ and $x_k \in \eta_i$). Figure \ref{fig:cov_matrix} shows the decomposition of the covariance matrix $\Sigma_{p}$ into the sub-matrix $\Sigma_{p}^{-}$ and the vector $\vec{\rho}$. By employing Kronecker products, we rewrite the first fundamental form (metric tensor) in a tensorial notation, providing a computationally efficient way to compute the elements of $g(\vec{\theta})$:

\begin{figure}[ht]
\begin{center}
\includegraphics[scale=0.4]{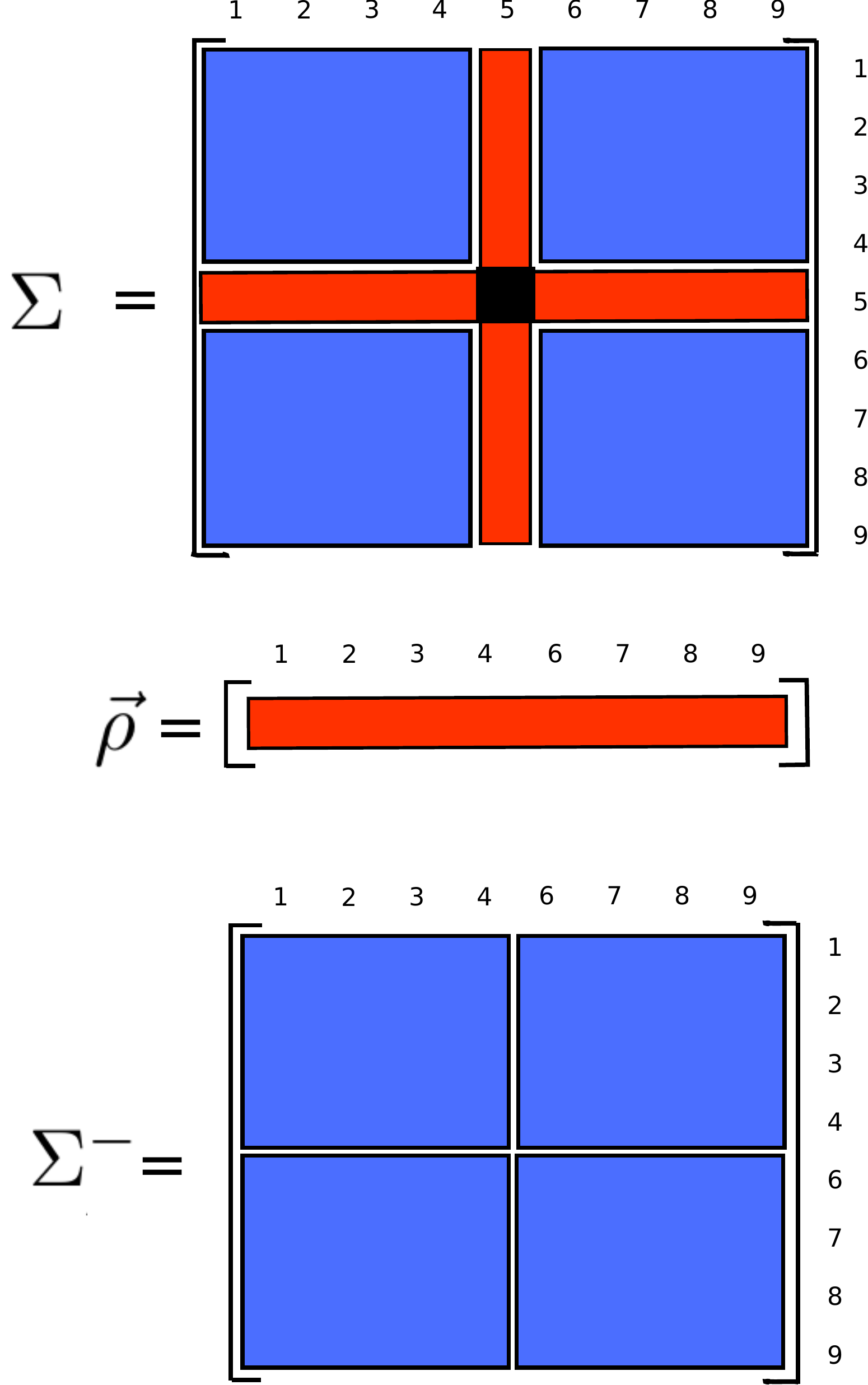}
\end{center}
\caption{Decomposition of $\Sigma_{p}$ into $\Sigma_{p}^{-}$ and $\vec{\rho}$ on a second-order neighborhood system ($\Delta=8$). By rewriting the components of the metric tensor in terms of Kronocker products, we can make numerical simulations faster.
}
\label{fig:cov_matrix}
\end{figure}

\begin{equation}
	g_{11} = \frac{1}{\sigma^2}\left(1-\beta \Delta \right)^2\left[ 1 - \frac{1}{\sigma^2}\left( 2\beta\left\| \vec{\rho} \right\|_{+} - \beta^2 \left\| \Sigma_{p}^{-} \right\|_{+} \right) \right] 
	\label{eq:g11}
\end{equation}

\begin{align}	
	\label{eq:g22}
	g_{22} = \frac{1}{2\sigma^4} & - \frac{1}{\sigma^6} \left[ 2\beta\left\| \vec{\rho} \right\|_{+} - \beta^2 \left\| \Sigma_{p}^{-} \right\|_{+} \right] \\ \nonumber & + \frac{1}{\sigma^8}\left[ 3\beta^2 \left\| \vec{\rho} \otimes \vec{\rho} \right\|_{+} - 3 \beta^3 \left\| \vec{\rho} \otimes \Sigma_{p}^{-} \right\|_{+} + 3\beta^4 \left\| \Sigma_{p}^{-} \otimes \Sigma_{p}^{-} \right\|_{+}  \right] \nonumber
\end{align}	
	
\begin{align}
	\label{eq:g23}
	g_{23} = g_{32} & = \frac{1}{\sigma^4}\left[ \left\| \vec{\rho} \right\|_{+} - \beta \left\| \Sigma_{p}^{-} \right\|_{+} \right] \\ \nonumber & - \frac{1}{2\sigma^6} \left[ 6\beta \left\| \vec{\rho} \otimes \vec{\rho} \right\|_{+} - 9 \beta^2 \left\| \vec{\rho} \otimes \Sigma_{p}^{-} \right\|_{+} + 3\beta^3 \left\| \Sigma_{p}^{-} \otimes \Sigma_{p}^{-} \right\|_{+}  \right] \nonumber
\end{align}	

\begin{equation}
	\label{eq:g33}
	g_{33} = \frac{1}{\sigma^2} \left\| \Sigma_{p}^{-} \right\|_{+} + \frac{1}{\sigma^4} \left[ 2 \left\| \vec{\rho} \otimes \vec{\rho} \right\|_{+} - 6 \beta \left\| \vec{\rho} \otimes \Sigma_{p}^{-} \right\|_{+} + 3\beta^2 \left\| \Sigma_{p}^{-} \otimes \Sigma_{p}^{-} \right\|_{+}  \right] 
\end{equation} where $\left\| A \right\|_{+}$ represents the summation of all the entries of the vector/matrix $A$ and $\otimes$ denotes the Kronecker (tensor) product. The advantage of this tensorial notation using Kronecker products is that we avoid the computation of triple and quadruple summations, which have complexity $O(n^3)$ and $O(n^4)$, respectively.

\subsection{Entropy in Gaussian random fields}

The concept of entropy is crucial in many areas of science. In summary, entropy is a thermodynamic quantity that measures the degree of molecular freedom of a system, and is associated with its number of configurations (or microstates), that is, how many ways particles can be distributed in quantized energy levels. It is also generally associated the randomness, dispersion of matter and energy, and ``disorder'' of a complex system. In random fields, it is possible to compute the entropy for each possible observed configuration. The entropy of a pairwise isotropic Gaussian-Markov random field can be computed by the expected value of self-information, which leads to:

\begin{align}
	\label{eq:entropia1}
	 H_{\beta}(\vec{\theta}) & = - E\left[ log~p\left(x_{i}| \eta_{i}, \vec{\theta} \right) \right] \\ \nonumber & = H_{G}(\vec{\theta}) - \frac{\beta}{\sigma^2}\sum_{j \in \eta_i}\sigma_{ij} + \frac{\beta^2}{2\sigma^2} \sum_{j \in \eta_i}\sum_{k \in \eta_i}\sigma_{jk}
\end{align} where $H_G(\vec{\theta})$ denotes the entropy of a Gaussian random variable. Note that the entropy is a quadratic function of the inverse temperature parameter $\beta$. Besides, for $\beta=0$, we have $H_{\beta}(\vec{\theta}) = H_{G}(\vec{\theta})$, as expected. Using the tensorial notation, the entropy can be expressed as: 

\begin{align}
	\label{eq:entropia2}
	 H_{\beta}(\vec{\theta}) = H_{G}(\vec{\theta}) - \frac{1}{\sigma^2} \left( \beta\left\| \vec{\rho} \right\|_{+} - \frac{\beta^2}{2} \left\| \Sigma_{p}^{-} \right\|_{+} \right)
\end{align}

\section{Geodesic distances in Gaussian random field manifolds}

In a Riemannian manifold $M$ equipped with a metric tensor $g$, the length of a curve $\gamma(t)$, defined for $t \in [a, b]$ is:

\begin{equation}
	L(\gamma) = \int_a^b \sqrt{g(\gamma'(t), \gamma'(t))} dt
\end{equation} where $\gamma'(t)$ denotes the fist derivative (tangent vector). It is easier to work with the energy of the curve. The energy of a curve $\gamma(t)$ is given by:

\begin{equation}
	E(\gamma) = \frac{1}{2} \int_a^b \sqrt{g(\gamma'(t), \gamma'(t))} dt
\end{equation} 

It is not difficult to show that if $M$ is a manifold equipped with a metric tensor $g(\theta)$, then:

\begin{equation}
	L(\gamma) \leq 2 (b - a) E(\gamma)
\end{equation} with equality if and only if $\gamma$ is parametrized by arc length, that is, if we move along the curve with constant speed (the tangent vectors are unitary), which means that $g(\gamma'(t), \gamma'(t))$ is constant. In conclusion a curve $\gamma(t)$ has minimum energy if and only if it has minimum length has arc length parametrization. A geodesic is a curve $\gamma(t)$ that minimizes the functional $L(\gamma)$. In order to obtain an exact solution, we need to solve the following optimization problem \cite{Riemann}:

\begin{equation}
	\nabla_{\gamma'} \gamma' = 0
\end{equation} where $\nabla$ is the covariant derivative. The geodesics are precisely the ``straight lines'' that have constant speed. From differential geometry, it has been shown that a solution to this equation must satisfy the geodesic equations, expressed by \cite{Tristan}:

\begin{equation}
	\frac{d^2 \theta_k}{dt^2} + \sum_{ij} \Gamma_{ij}^k \frac{d \theta_i}{dt} \frac{d \theta_j}{dt} = 0
\end{equation} where, in our pairwise isotropic Gaussian-Markov random field, $\theta_1 = \mu$, $\theta_2 = \sigma^2$ and $\theta_3 = \beta$ are the model parameters and $\Gamma_{ij}^k$ are the Christoffel symbols of the metric, defined as \cite{Manfredo}:

\begin{equation}
	\Gamma_{ij}^k = \frac{1}{2}\sum_{m=1}^{3} \left( \frac{\partial}{\partial\theta_i}g_{jm} + \frac{\partial}{\partial\theta_j}g_{im} - \frac{\partial}{\partial\theta_m}g_{ij} \right) g^{mk}
\end{equation} where $g_{ij}$ a component of the metric tensor (Fisher information matrix) and $g^{ij}$ is a component of the inverse metric tensor. As the parametric space of pairwise isotropic Gaussian-Markov random fields are 3D manifolds, we have a total of $3 \times 3 \times 3 = 27$ Christoffel symbols arranged in three $3 \times 3$ matrices $\Gamma^1$, $\Gamma^2$ and $\Gamma^3$, leading to the following system of non-linear differential equations:

\begin{align}
\theta_1''(t) - \sum_{ij} \Gamma_{ij}^{1} \theta_i'(t) \theta_j'(t) = 0 \\
\theta_2''(t) - \sum_{ij} \Gamma_{ij}^{2} \theta_i'(t) \theta_j'(t) = 0 \\
\theta_3''(t) - \sum_{ij} \Gamma_{ij}^{3} \theta_i'(t) \theta_j'(t) = 0
\end{align} where $\theta_i'(t)$ and $\theta_i''(t)$ denotes the first and second derivatives of the coordinates, the values of $\theta_1^{(0)} = \mu_0$, $\theta_2^{(0)} = \sigma_0^2$ and $\theta_3^{(0)} = \beta_0$ represent the starting point A, and the values of $\theta_1'^{(0)}$, $\theta_2'^{(0)}$ and $\theta_3'^{(0)}$ defines the initial tangent vector at A (it points to the initial direction that we want move along the space). The fourth-order Runge-Kutta method can be applied to obtain a numerical solution for this system of differential equations, where the final value of the parameter $t$ (simulation time and number of iterations) determines how long the geodesic curve will grow. Note that there are no explicit values for the final parameters $\theta_1^{(t)} = \mu_t$, $\theta_2^{(t)} = \sigma_t^2$ and $\theta_3^{(t)} = \beta_t$, since the manifold is a curved space and the geodesic is build locally, using one tangent space at a time. 

\subsection{The Christoffel symbols of Gaussian random field manifolds}

In order to compute the Christoffel symbols of the underlying parametric space of pairwise isotropic Gaussian-Markov random field models, we need to find the components of the inverse metric tensor $g^{-1}(\theta)$:

\begin{equation}
	g^{-1}(\vec{\theta}) = \left( \begin{array}{ccc}
	g^{11} & 0 & 0 \\ 
	0 & g^{22} & g^{23} \\ 
	0 & g^{32} & g^{33}
	\end{array} \right)
\end{equation} 

It is easy to see that the components of the inverse metric tensor are:

\begin{align}
	g^{11} & = \frac{1}{g_{11}} \\ 
	g^{22} & = \frac{g_{33}}{g_{22}g_{33} - g_{23}^2} \\ 
	g^{23} & = \frac{g_{23}}{g_{23}^2 - g_{22}g_{33}} \\
	g^{32} & = \frac{g_{32}}{g_{32}^2 - g_{22}g_{33}} \\
	g^{33} & = \frac{g_{22}}{g_{22}g_{33} - g_{32}^2}
\end{align} where $g^{23} = g^{32}$. In the computational experiments, we perform regularization to avoid numerical instability by adding a small value $\lambda = 0.01$ to the main diagonal of the metric tensor. The next step consists in the derivation of the components of the Christoffel symbols. First, note that:

\begin{equation}
	\frac{\partial}{\partial\theta_1} g_{ij} = \frac{\partial}{\partial\mu} g_{ij} = 0, \qquad\qquad \forall i, j
\end{equation} since none of the components of the metric tensor is a function of $\theta_1 = \mu$. Moving to the derivatives with respect to $\theta_2 = \sigma^2$, we have:

\begin{align}
	\frac{\partial}{\partial\theta_2}g_{11} & = -\frac{1}{\sigma^4}(1 - \beta\Delta)^2 + \frac{2}{\sigma^6}(1 - \beta\Delta)^2 \left( 2\beta \left\| \vec{\rho} \right\|_{+} - \beta^2 \left\| \Sigma_{p}^{-} \right\|_{+} \right) \\
	\frac{\partial}{\partial\theta_2}g_{22} & = -\frac{1}{\sigma^6} + \frac{3}{\sigma^8} \left( 2\beta \left\| \vec{\rho} \right\|_{+} - \beta^2 \left\| \Sigma_{p}^{-} \right\|_{+} \right) \\ \nonumber & \qquad - \frac{4}{\sigma^{10}} \left( 3\beta^2 \left\| \vec{\rho} \otimes \vec{\rho} \right\|_{+} - 3 \beta^3 \left\| \vec{\rho} \otimes \Sigma_{p}^{-} \right\|_{+} + 3 \beta^4  \left\| \Sigma_{p}^{-} \otimes \Sigma_{p}^{-} \right\|_{+} \right) \\
	\frac{\partial}{\partial\theta_2}g_{23} & = \frac{\partial}{\partial\theta_2}g_{32} = -\frac{2}{\sigma^6}\left( \left\| \vec{\rho} \right\|_{+} - \beta \left\| \Sigma_{p}^{-} \right\|_{+} \right) \\ \nonumber & \qquad + \frac{3}{2\sigma^8}\left( 6\beta \left\| \vec{\rho} \otimes \vec{\rho} \right\|_{+} - 9\beta^2 \left\| \vec{\rho} \otimes \Sigma_{p}^{-} \right\|_{+} + 3 \beta^3 \left\| \Sigma_{p}^{-} \otimes \Sigma_{p}^{-} \right\|_{+} \right) \\
	\frac{\partial}{\partial\theta_2}g_{33} & = -\frac{1}{\sigma^4} \left\| \Sigma_{p}^{-} \right\|_{+} \\ \nonumber & \qquad - \frac{2}{\sigma^6}\left( 2 \left\| \vec{\rho} \otimes \vec{\rho} \right\|_{+} - 6\beta \left\| \vec{\rho} \otimes \Sigma_{p}^{-} \right\|_{+} + 3 \beta^2 \left\| \Sigma_{p}^{-} \otimes \Sigma_{p}^{-} \right\|_{+} \right)
\end{align}

Finally, we have to differentiate the components of the metric tensor with respect to the inverse temperature parameter, that is, $\theta_3 = \beta$:


\begin{align}
	\frac{\partial}{\partial\theta_3}g_{11} & = -\frac{1}{\sigma^2}2\Delta(1 - \beta\Delta) \left[ 1 - \frac{1}{\sigma^2}\left( 2\beta\left\| \vec{\rho} \right\|_{+} - \beta^2 \left\| \Sigma_{p}^{-} \right\|_{+} \right) \right] \\ \nonumber & \qquad\qquad\qquad\qquad - \frac{1}{\sigma^4} (1 - \beta\Delta)^2 \left( 2\left\| \vec{\rho} \right\|_{+} - 2\beta\left\| \Sigma_{p}^{-} \right\|_{+} \right) \\
	\frac{\partial}{\partial\theta_3}g_{22} & = -\frac{1}{\sigma^6}\left( 2\left\| \vec{\rho} \right\|_{+} - 2\beta \left\| \Sigma_{p}^{-} \right\|_{+} \right) \\ \nonumber & \qquad + \frac{1}{\sigma^8} \left( 6\beta \left\| \vec{\rho} \otimes \vec{\rho} \right\|_{+} - 9 \beta^2 \left\| \vec{\rho} \otimes \Sigma_{p}^{-} \right\|_{+} + 12 \beta^3  \left\| \Sigma_{p}^{-} \otimes \Sigma_{p}^{-} \right\|_{+} \right) \\
	\frac{\partial}{\partial\theta_3}g_{23} & = \frac{\partial}{\partial\theta_3}g_{32} = -\frac{1}{\sigma^4}\left\| \Sigma_{p}^{-} \right\|_{+} \\ \nonumber & \qquad - \frac{1}{2\sigma^6}\left( 6\left\| \vec{\rho} \otimes \vec{\rho} \right\|_{+} - 18\beta \left\| \vec{\rho} \otimes \Sigma_{p}^{-} \right\|_{+} + 9 \beta^2 \left\| \Sigma_{p}^{-} \otimes \Sigma_{p}^{-} \right\|_{+} \right) \\
	\frac{\partial}{\partial\theta_3}g_{33} & = -\frac{1}{\sigma^4} \left( 6\beta \left\| \vec{\rho} \otimes \Sigma_{p}^{-} \right\|_{+} - 6 \beta \left\| \Sigma_{p}^{-} \otimes \Sigma_{p}^{-} \right\|_{+} \right)
\end{align}

Given the above, it is possible to compute the 27 Christoffel symbols, denoted by: $\Gamma_{11}^1, \Gamma_{11}^2, \Gamma_{11}^3, \Gamma_{12}^1, \Gamma_{12}^2, \Gamma_{12}^3, ..., \Gamma_{33}^1, \Gamma_{33}^2, \Gamma_{33}^3$, organized in a tensor composed by $\Gamma^1, \Gamma^2, \Gamma^3$. However, it can be shown that the Christoffel symbols are symmetric with respect to the lower indices, that is, $\Gamma_{ij}^k = \Gamma_{ji}^k$, resulting in a total of 18 different symbols. First, note that the components $\Gamma_{11}^1, \Gamma_{11}^2, \Gamma_{11}^3$ are:


\begin{align}
	\Gamma_{11}^1 & = 0 \\
	\Gamma_{11}^2 & = -\frac{1}{2}\left( \frac{\partial g_{11}}{\partial\theta_2}g^{22} + \frac{\partial g_{11}}{\partial\theta_3}g^{32} \right) \\
	\Gamma_{11}^3 & = -\frac{1}{2}\left( \frac{\partial g_{11}}{\partial\theta_2}g^{23} + \frac{\partial g_{11}}{\partial\theta_3}g^{33} \right)
\end{align}

Next, the components $\Gamma_{12}^1, \Gamma_{12}^2, \Gamma_{12}^3$ can be expressed by:



\begin{align}
	\Gamma_{12}^1 = \frac{1}{2} \frac{\partial g_{11}}{\partial\theta_2}g^{11} = \Gamma_{21}^1 \qquad
	\Gamma_{12}^2 = \Gamma_{21}^2 = 0 \qquad
	\Gamma_{12}^3 = \Gamma_{21}^3 = 0
\end{align}

Similarly, the components $\Gamma_{13}^1, \Gamma_{13}^2, \Gamma_{13}^3$ also have the same structure:

\begin{align}
	\Gamma_{13}^1 = \frac{1}{2} \frac{\partial g_{11}}{\partial\theta_3}g^{11} = \Gamma_{31}^1 \qquad
	\Gamma_{13}^2 = \Gamma_{31}^2 = 0 \qquad
	\Gamma_{13}^3 = \Gamma_{31}^3 = 0
\end{align}

Moving forward to the components $\Gamma_{22}^1, \Gamma_{22}^2, \Gamma_{22}^3$, we have:


\begin{align}
	\Gamma_{22}^1 & = 0 \\
	\Gamma_{22}^2 & = \frac{1}{2} \left[ \frac{\partial g_{22}}{\partial\theta_2}g^{22} + \left( 2 \frac{\partial g_{23}}{\partial\theta_2} - \frac{\partial g_{22}}{\partial\theta_3} \right)g^{32} \right] \\
	\Gamma_{22}^3 & = \frac{1}{2} \left[ \frac{\partial g_{22}}{\partial\theta_2}g^{23} + \left( 2 \frac{\partial g_{23}}{\partial\theta_2} - \frac{\partial g_{22}}{\partial\theta_3} \right)g^{33} \right]
\end{align}

The components $\Gamma_{23}^1, \Gamma_{23}^2, \Gamma_{23}^3$ are given by:


\begin{align}
	\Gamma_{23}^1 & = \Gamma_{32}^1 = 0 \\
	\Gamma_{23}^2 & = \frac{1}{2}\left( \frac{\partial g_{22}}{\partial\theta_3}g^{22} + \frac{\partial g_{33}}{\partial\theta_2}g^{32} \right) = \Gamma_{32}^2 \\
	\Gamma_{23}^3 & = \frac{1}{2}\left( \frac{\partial g_{22}}{\partial\theta_3}g^{23} + \frac{\partial g_{33}}{\partial\theta_2}g^{33} \right) = \Gamma_{32}^3
\end{align}

Finally, the components $\Gamma_{33}^1, \Gamma_{33}^2, \Gamma_{33}^3$ can be computed by:


\begin{align}
	\Gamma_{33}^1 & = 0 \\
	\Gamma_{33}^2 & = \frac{1}{2} \left[ \left( 2 \frac{\partial g_{32}}{\partial\theta_3} - \frac{\partial g_{33}}{\partial\theta_2} \right) g^{22} + \frac{\partial g_{33}}{\partial\theta_3}g^{32} \right] \\
	\Gamma_{33}^3 & = \frac{1}{2} \left[ \left( 2 \frac{\partial g_{32}}{\partial\theta_3} - \frac{\partial g_{33}}{\partial\theta_2} \right) g^{23} + \frac{\partial g_{33}}{\partial\theta_3}g^{33} \right] 
\end{align}

Therefore, from the 27 Christoffel symbols, 13 are zero and the remaining 14 non-zero symbols assume 10 different values. In order to numerically solve our system of non-linear second-order differential equations, first we have to convert it to a first-order system. This can be done by a simple variable substitution. Let $\gamma_1(t) = \theta_1(t)$ and $\alpha_1(t) = \theta_1'(t)$. Then, we have:

\begin{align}
	\gamma_1'(t) & = \theta_1'(t) = \alpha_1(t) \\
	\alpha_1'(t) & = \theta_1''(t) = - \sum_{ij} \Gamma_{i,j}^1 \alpha_i(t) \alpha_j(t)
\end{align}

By direct application of this procedure to the other equations, we finally reach our system of non-linear first-order differential equations:

\begin{align}
	\gamma_1'(t) & = \alpha_1(t) \\
	\gamma_2'(t) & = \alpha_2(t) \\
	\gamma_3'(t) & = \alpha_3(t) \\
	\alpha_1'(t) & = - \sum_{ij} \Gamma_{i,j}^1 \alpha_i(t) \alpha_j(t)\\
	\alpha_2'(t) & = - \sum_{ij} \Gamma_{i,j}^2 \alpha_i(t) \alpha_j(t)\\
	\alpha_3'(t) & = - \sum_{ij} \Gamma_{i,j}^3 \alpha_i(t) \alpha_j(t)
\end{align}

Note that we can rewrite our system in a more compact way as:

\begin{align}
	\gamma_1'(t) & = F(t, \vec{\alpha}) = \alpha_1 \\
	\gamma_2'(t) & = G(t, \vec{\alpha}) = \alpha_2 \\
	\gamma_3'(t) & = H(t, \vec{\alpha}) = \alpha_3 \\
	\alpha_1'(t) & = P(t, \vec{\alpha}, \Gamma_1) = - \vec{\alpha}^T \Gamma^1 \vec{\alpha} \\
	\alpha_2'(t) & = Q(t, \vec{\alpha}, \Gamma_2) = - \vec{\alpha}^T \Gamma^2 \vec{\alpha} \\
	\alpha_3'(t) & = R(t, \vec{\alpha}, \Gamma_3) = - \vec{\alpha}^T \Gamma^3 \vec{\alpha}
\end{align} where $\vec{\alpha}^T = [ \alpha_1, \alpha_2, \alpha_3 ]$ is the tangent vector and:

\begin{equation}
	\Gamma^k = \left( \begin{array}{ccc}
	\Gamma_{11}^k & \Gamma_{12}^k & \Gamma_{13}^k \\ 
	\Gamma_{21}^k & \Gamma_{22}^k & \Gamma_{23}^k \\ 
	\Gamma_{31}^k & \Gamma_{32}^k & \Gamma_{33}^k \\ 
	\end{array} \right)
\end{equation}

\begin{algorithm}[H]
\caption{Computing the geodesic distance with 4th order Runge-Kutta}\label{alg:gd}
\begin{algorithmic}[1]
\Function{GeodesicDistance}{$a$, $b$, $n$, $\vec{\gamma}^{(0)}$, $\vec{\alpha}^{(0)}$}
\State $h = (b - a)/n$
\State $t = a$
\State $dist = 0$
\For{$i = 0$, $i < MAX$, $i++$} 
	\State Generate an outcome of the GMRF model using $\vec{\gamma}^{(i)}$
	\State Compute the metric tensor $g(\vec{\gamma}^{(i)})$ with equations \ref{eq:g11}, \ref{eq:g22}, \ref{eq:g23} and \ref{eq:g33}
	\State Compute the Christoffel symbols ($\Gamma_1$, $\Gamma_2$, $\Gamma_3$) with equations 68-81
	\State Perform a 4th order Runge Kutta iteration as
	\State $k_0 = hF(t, \vec{\alpha})$, $l_0 = hG(t, \vec{\alpha})$, $m_0 = hH(t, \vec{\alpha})$			
	\State $x_0 = hP(t, \vec{\alpha})$,	$y_0 = hQ(t, \vec{\alpha})$, $z_0 = hR(t, \vec{\alpha})$	
	\State $\vec{\delta} = [k_0, l_0, m_0]$
	\State $k_1 = hF(t+0.5h, \vec{\alpha}+0.5\vec{\delta})$
	\State $l_1 = hG(t+0.5h, \vec{\alpha}+0.5\vec{\delta})$
	\State $m_1 = hH(t+0.5h, \vec{\alpha}+0.5\vec{\delta})$			
	\State $x_1 = hP(t+0.5h, \vec{\alpha}+0.5\vec{\delta}, \Gamma_1)$
	\State $y_1 = hQ(t+0.5h, \vec{\alpha}+0.5\vec{\delta}, \Gamma_2)$
	\State $z_1 = hR(t+0.5h, \vec{\alpha}+0.5\vec{\delta}, \Gamma_3)$
	\State $\vec{\delta} = [k_1, l_1, m_1]$
	\State $k_2 = hF(t+0.5h, \vec{\alpha}+0.5\vec{\delta})$
	\State $l_2 = hG(t+0.5h, \vec{\alpha}+0.5\vec{\delta})$
	\State $m_2 = hH(t+0.5h, \vec{\alpha}+0.5\vec{\delta})$			
	\State $x_2 = hP(t+0.5h, \vec{\alpha}+0.5\vec{\delta}, \Gamma_1)$
	\State $y_2 = hQ(t+0.5h, \vec{\alpha}+0.5\vec{\delta}, \Gamma_2)$
	\State $z_2 = hR(t+0.5h, \vec{\alpha}+0.5\vec{\delta}, \Gamma_3)$
	\State $\vec{\delta} = [k_2, l_2, m_2]$
	\State $k_2 = hF(t+h, \vec{\alpha}+\vec{\delta})$
	\State $l_2 = hG(t+h, \vec{\alpha}+\vec{\delta})$
	\State $m_2 = hH(t+h, \vec{\alpha}+\vec{\delta})$			
	\State $x_2 = hP(t+h, \vec{\alpha}+\vec{\delta}, \Gamma_1)$
	\State $y_2 = hQ(t+h, \vec{\alpha}+\vec{\delta}, \Gamma_2)$
	\State $z_2 = hR(t+h, \vec{\alpha}+\vec{\delta}, \Gamma_3)$
	\State Update the vectors $\vec{\gamma}^{(i)}$ (position) and $\vec{\alpha}^{(i)}$ (speed) as
	\State $\gamma_1^{(i+1)} = \gamma_1^{(i)} + (1/6)(k_0 + 2k_1 + 2k_2 + k_3)$
	\State $\gamma_2^{(i+1)} = \gamma_2^{(i)} + (1/6)(l_0 + 2l_1 + 2l_2 + l_3)$
	\State $\gamma_3^{(i+1)} = \gamma_3^{(i)} + (1/6)(m_0 + 2m_1 + 2m_2 + m_3)$
	\State $\alpha_1^{(i+1)} = \alpha_1^{(i)} + (1/6)(x_0 + 2x_1 + 2x_2 + x_3)$
	\State $\alpha_2^{(i+1)} = \alpha_2^{(i)} + (1/6)(y_0 + 2y_1 + 2y_2 + y_3)$
	\State $\alpha_3^{(i+1)} = \alpha_3^{(i)} + (1/6)(z_0 + 2z_1 + 2z_2 + z_3)$
	\State Update the geodesic distance: $dist = dist + \lVert \alpha(t) \rVert h$
\EndFor
\EndFunction
\end{algorithmic}
\end{algorithm}

\begin{figure}[ht]
\begin{center}
\centerline{\includegraphics[scale=0.6]{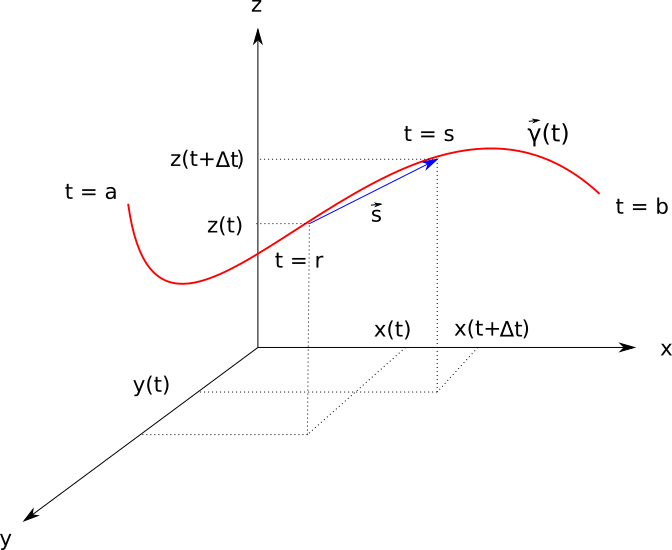}}
\caption{The computation of the arc length can be approximated by the norm of the difference vector (hypotenuse of a right triangle).}
\label{fig:arc_length}
\end{center}
\end{figure}

\begin{table}[]
\centering
\small
\caption{Comparison between geodesic and Euclidean distances for several pairwise isotropic Gaussian-Markov random field models}
\begin{tabular}{ccccccccccc}
\toprule
\multicolumn{3}{c}{\textbf{Initial position}}                    & \multicolumn{3}{c}{\textbf{Tangent vector}}                    & \multicolumn{3}{c}{\textbf{Final position}}  \\
\midrule
$\mu_a$ & $\sigma_a^2$ & $\beta_a$ & $\alpha_1$ & $\alpha_2$ & $\alpha_3$ & $\mu_b$ & $\sigma_b^2$ & $\beta_b$ & \textbf{G.D.} & \textbf{E.D.} \\
\midrule
0.0 & 1.0 & 0.0 & 0.0 & 0.0 & 0.1 & 0.0 & 1.203 & 0.465 & 0.686 & 0.631 \\
0.0 & 1.0 & 0.0 & 0.1 & 0.1 & 0.2 & 0.921 & 1.116 & -0.655 & 1.667 & 1.137 \\
5.0 & 10.0 & 0.0 & 0.1 & 0.1 & -0.1 & 5.102 & 11.525 & -0.487 & 1.629 & 1.596 \\
5.0 & 10.0 & -1.0 & 0.1 & -0.1 & 0.2 & 7.238 & 12.495 & -1.379 & 3.794 & 3.37 \\
5.0 & 10.0 & 0.5 & -0.1 & -0.1 & -0.2 & 1.908 & 12.031 & 1.451 & 5.292 & 3.819 \\
1.0 & 1.0 & -1.0 & 0.2 & 0.2 & 0.2 & 1.568 & 1.720 & -2.736 & 2.194 & 1.933 \\
0.0 & 100.0 & 0.0 & 0.2 & -1.0 & 0.2 & 0.586 & 96.500 & 0.600 & 3.58 & 3.53 \\
1.0 & 1.0 & -1.0 & 0.02 & 0.02 & 0.2 & 1.705 & 2.550 & -1.786 & 2.863 & 1.879 \\
1.0 & 1.0 & 0.0 & 0.05 & 0.05 & 0.05 & 1.681 & 1.141 & -0.130 & 0.805 & 0.702 \\
1.0 & 1.0 & 0.0 & -0.05 & -0.05 & 0.1 & 0.212 & 0.581 & -0.230 & 1.161 & 0.905 \\
10.0 & 5.0 & 0.0 & -0.25 & 0.25 & 0.8 & 9.818 & 6.084 & 1.311 & 1.942 & 1.662 \\
10.0 & 5.0 & 0.0 & 2.0 & 0.05 & 0.1 & 10.891 & 5.160 & -1.133 & 1.819 & 1.383 \\
5.0 & 1.0 & 0.0 & 0.0 & 0.5 & 0.2 & 5.0 & 2.169 & 0.879 & 1.556 & 1.431 \\
0.0 & 1.0 & 0.0 & 0.01 & 0.5 & 0.2 & 0.068 & 2.194 & 0.893 & 1.579 & 1.461 \\
5.0 & 10.0 & -0.5 & 0.01 & 0.01 & 0.2 & 5.863 & 13.438 & 0.855 &  4.503 & 4.027 \\
\bottomrule
\end{tabular}
\label{tab:ds}
\end{table}

The basic idea is that the initial condition $\vec{\gamma}^{(0)} = [ \gamma_1^{(0)}, \gamma_2^{(0)}, \gamma_3^{(0)} ]$ defines the starting point in the manifold and the initial condition $\vec{\alpha}^{(0)} = [ \alpha_1^{(0)}, \alpha_2^{(0)}, \alpha_3^{(0)} ]$ defines the initial tangent vector (speed). The intuition behind this process is as follows: suppose we place a marble in the manifold. At each iteration, the system of differential equations updates the position and the speed of the marble, based in the initial conditions. With these equations, we can completely characterize the trajectory of the moving marble along geodesics in the manifold. A limitation, however, is that, in order to compute the Christoffel symbols, we need to estimate the covariance matrix of the patches along the random field. In order to do so, we use Markov Chain Monte Carlo simulation to generate an outcome of the pairwise isotropic Gaussian-Markov random field at each iteration of the Runge-Kutta method using the parameter vector $\vec{\gamma}$, which is nothing more than the position of the marble in the manifold. Hence, as the marble moves along the manifold, we have different outcomes for our random field model. The arc length of the resulting geodesic curve defines our geodesic distance. Figure \ref{fig:arc_length} illustrates how the arc length between $t = r$ and $t = s$ can be approximated by the norm of the vector $\vec{s}$, which is the difference between $\vec{\gamma}(r)$ and $\vec{\gamma}(s)$. If we increase the number of subdivisions of the interval $[r, s]$, the better will be the approximation, given by:

\begin{equation}
	s(t) \approx \sum_{t=r}^{s} \sqrt{ \left( x(t+\Delta t) - x(t) \right)^2 + \left( y(t+\Delta t) - y(t) \right)^2 + \left( z(t+\Delta t) - z(t) \right)^2 }
\end{equation}

Note that  we can rewrite $s(t)$ as:

\begin{equation}
	s(t) \approx \sum_{t=r}^{s} \sqrt{ \left( \frac{x(t+\Delta t) - x(t)}{\Delta t} \right)^2 + \left( \frac{y(t+\Delta t) - y(t)}{\Delta t} \right)^2 + \left( \frac{z(t+\Delta t) - z(t)}{\Delta t} \right)^2 } \Delta t
\end{equation}

From the definition of the derivative of a function, we have:

\begin{equation}
	f'(t) = \lim\limits_{\Delta t \to 0} \frac{f(t+\Delta t) - f(t)}{\Delta t}
\end{equation}

\begin{figure}[!ht]
	\begin{center}
	\includegraphics[scale=0.37]{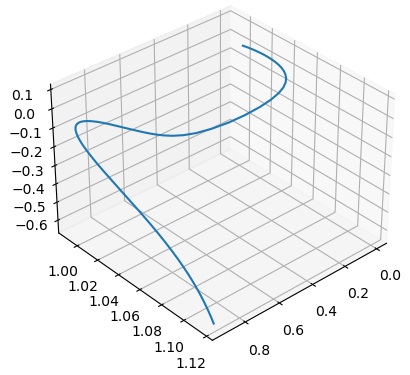}
	\includegraphics[scale=0.37]{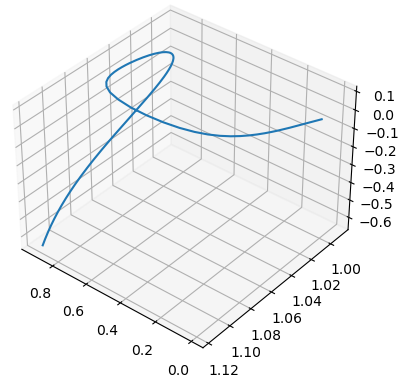}
	\includegraphics[scale=0.37]{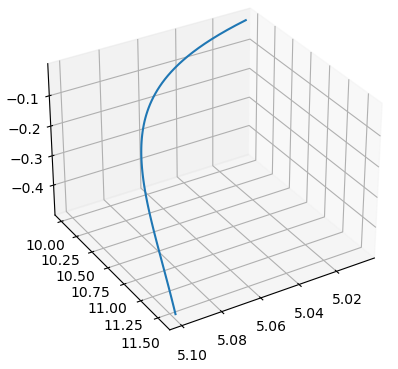}
	\includegraphics[scale=0.37]{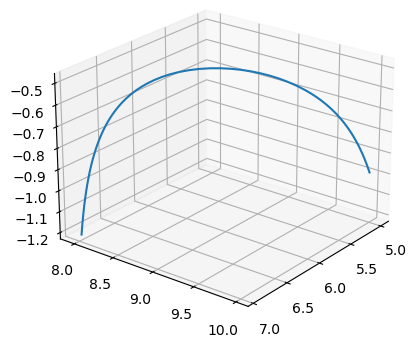}
	\end{center}
	\caption{Some geodesic curves obtained from the numerical simulations.}
	\label{fig:geodesic1}
\end{figure}

\begin{figure}[!ht]
	\begin{center}
	\includegraphics[scale=0.37]{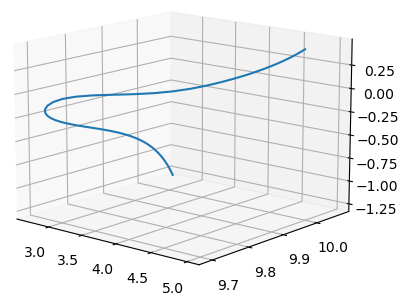}
	\includegraphics[scale=0.37]{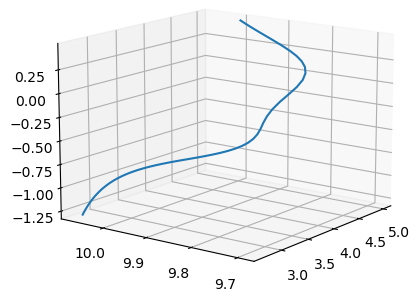}
	\includegraphics[scale=0.37]{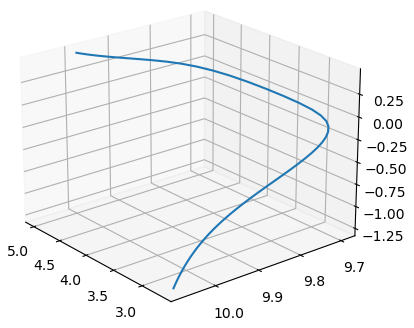}
	\includegraphics[scale=0.37]{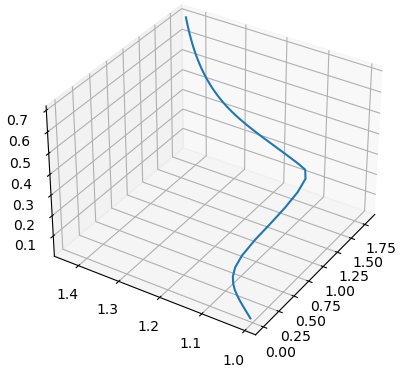}
	\end{center}
	\caption{More examples of geodesic curves obtained from the numerical simulations.}
	\label{fig:geodesic2}
\end{figure}

Thus, in the limiting case, when $\Delta t \to 0$, the approximation becomes exact:

\begin{align}
	s(t) & = \int_{a}^{b} \sqrt{ x'(t)^2 + y'(t)^2 + z'(t)^2 } dt \\ \nonumber & = \int_{a}^{b} \sqrt{ \vec{\gamma}'(t)^T \vec{\gamma}'(t)  } dt = \int_{a}^{b} \lVert \vec{\gamma}'(t) \rVert dt
\end{align} where the integrand $\lVert \vec{\gamma}'(t) \rVert$ is the norm of the tangent vector at $t$. Therefore, we compute the geodesic distance as:


\begin{equation}
	L(\gamma) = \int_{a}^{b}  \lVert \gamma'(t) \rVert dt \approx \sum_{i=a}^{b} \lVert \gamma'(i) \rVert   h
\end{equation} using a small enough step size $h=(b-a)/n$. In our computational experiments, we set $a = 0$, $b = 5$, $n = 200$ and vary the initial conditions $\vec{\gamma}^{(0)}$ (starting point) and $\vec{\alpha}^{(0)}$ (initial speed). Table \ref{tab:ds} shows the initial conditions, final point and obtained geodesic distances for several executions of the proposed algorithm. Figures \ref{fig:geodesic1} and \ref{fig:geodesic2} illustrate the representation in $R^3$ for some of the geodesic curves obtained in the computational experiments.

It is worth to mention that the initial conditions must be defined in a way that the magnitude of the initial tangent vector (initial speed) is not too large. In practical terms, we have observed in our simulations that the system of differential equations exhibit a stable behavior if the components of the vector $\vec{\alpha}^{(0)}$ belong to the interval $[-0.5, 0.5]$. However, a theoretical analysis of the stability of the proposed system of differential equations is required to achieve mathematical guarantees in terms of convergence. 

\begin{figure}[!ht]
	\begin{center}
	\includegraphics[scale=0.37]{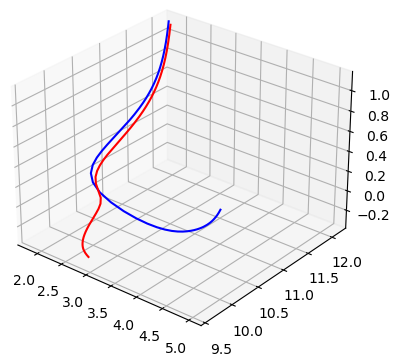}
	\includegraphics[scale=0.37]{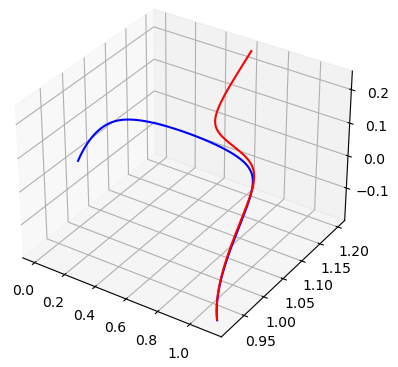}
	\includegraphics[scale=0.37]{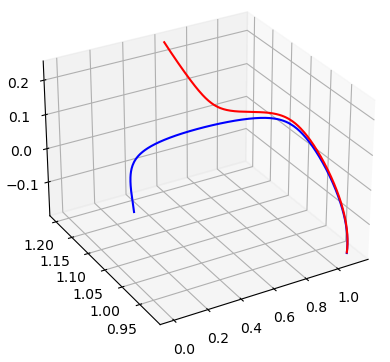}
	\includegraphics[scale=0.37]{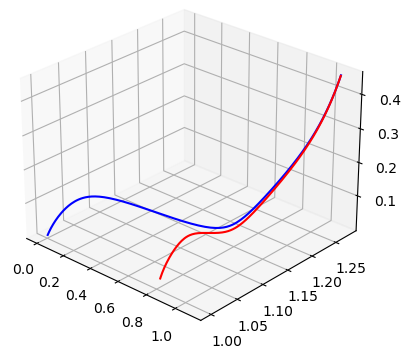}
	\end{center}
	\caption{Example of geodesic curves and their respective time reversed versions}
	\label{fig:reverse}
\end{figure}

One interesting observation concerns the time reversal of the numerical simulation. Suppose we start the simulation in a point A in the manifold and after $n$ steps we reach the ending point B with a final tangent vector $\vec{\alpha}(B)$ through a geodesic curve $\vec{\gamma}(t)$, for $t \in [A, B]$. Next, we start the simulation in the point B using the same tangent vector $\vec{\alpha}(B)$ in time reverse. Our initial expectation is that we would recover the same trajectory in reverse order, but the results show that at some point there is a bifurcation, that is, it is not possible to recover the exact curve. Figure \ref{fig:reverse} shows some of the geodesic curves, where the blue one is the original and the red one is the time reversed version. This behavior clearly shows that in Gaussian random field manifolds, the geodesic curve from the point A to the point B can be different from the geodesic curve from the point B to the point A, suggesting the existence of irreversible geometric deformations when the system undergoes phase transitions.

For the interested reader, the complete Python source code used in the computational experiments described in this paper is available at Github in the following link: \url{https://github.com/alexandrelevada/RK4_Geodesic_GMRF}. 

\section{Conclusions and final remarks}

Geodesic curves are like straight lines between two points in a manifold, because they minimize the distance (arc length) between them. Several works about the computation of geodesic distances between independent random variables have been proposed in the information geometry literature. However, little is known about situations in which a spatial dependence structure emerge in time. 

In this paper, we proposed a computational method to build geodesic curves in Gaussian random field manifolds. In order to simplify the mathematical derivations two assumptions were considered: the existence of pairwise interactions only, the hypothesis that the inverse temperature parameter is spatially invariant and isotropic. Basically, to achieve our goal, three major steps were required: 1) the derivation of the metric tensor of the underlying parametric spaces of pairwise isotropic Gaussian-Markov random fields (first-order Fisher information matrix); 2) the derivation of the Christoffel symbols, mathematical structures that are useful in defining how an object moves along the manifold; and 3) the development of a numerical algorithm based in the fourth-order Runge Kutta method and Markov-Chain Monte Carlo simulation to solve the system of non-linear differential equations that define the trajectory along geodesic curves in the manifold.

The obtained results show that in some cases the difference between the Euclidean distance and the geodesic distance is small, and the former can be considered a reasonable approximation for the later. However, there are situations in which the curvatures along the path make the geodesic distance almost 40\% larger than the Euclidean distance. Moreover, an interesting observation concerns the time reversal geodesic curves. When we run the simulation backwards in time, it is not possible to travel along the original geodesic curve, showing a bifurcation pattern. We still do not have a definitive answer to why this behavior is observed in our computational experiments, but one possible explanation would be the presence of irreversible geometric deformations in the trajectory when a moving reference travels along a geodesic curve. 

Future works may include the mathematical derivation of the Riemann curvature tensor of Gaussian random field manifolds in order to compute the Gaussian curvature of the manifold using the intrinsic geometry. We also intend to use the proposed method to replace the Euclidean distance in data science and machine learning applications. Finally, we intend to apply the proposed method to compute geodesic curves between other random field models besides Gaussian random fields, such as the Ising and the q-state Potts model, as well as other statistical models that do not belong to the exponential family.


\bibliography{mybibfile}

\begin{thebibliography}{10}
\expandafter\ifx\csname url\endcsname\relax
  \def\url#1{\texttt{#1}}\fi
\expandafter\ifx\csname urlprefix\endcsname\relax\def\urlprefix{URL }\fi
\expandafter\ifx\csname href\endcsname\relax
  \def\href#1#2{#2} \def\path#1{#1}\fi

\bibitem{Graph_dist}
Y.~Shimada, Y.~Hirata, T.~Ikeguchi, K.~Aihara, Graph distance for complex
  networks, Scientific Reports 6~(1) (2016) 449--455.

\bibitem{Frontiers}
J.~G. Diaz~Ochoa, Observability of complex systems by means of relative
  distances between homological groups, Frontiers in Physics 8 (2020) 503.

\bibitem{Entropy}
C.~Shea-Blymyer, S.~Roy, B.~Jantzen, A general metric for the similarity of
  both stochastic and deterministic system dynamics, Entropy 23~(9).

\bibitem{similarity}
A.~Charfi, S.~Ammar~Bouhamed, E.~Bosse, I.~Khanfir~Kallel, W.~Bouchaala,
  B.~Solaiman, N.~Derbel, Possibilistic similarity measures for data science
  and machine learning applications, IEEE Access 8 (2020) 49198--49211.

\bibitem{Physical_Systems}
M.~Concoyle, G.~Coatmundi, The Mathematical Structure of Stable Physical
  Systems, Trafford Publishing, 2014.

\bibitem{Tristan}
T.~Needham, Visual Differential Geometry and Forms: A Mathematical Drama in
  Five Acts, Princeton University Press, 2021.

\bibitem{Amari2021}
S.~Amari, Information geometry, International Statistical Review 89~(2) (2021)
  250--273.

\bibitem{Dodson}
K.~Arwini, C.~T.~J. Dodson, Information Geometry: Near Randomness and Near
  Independence, Springer, 2008.

\bibitem{Pinele}
J.~Pinele, J.~E. Strapasson, S.~I.~R. Costa, The fisher–rao distance between
  multivariate normal distributions: Special cases, boundsand applications,
  Entropy 22~(4) (2020) 404.

\bibitem{GaussianRandomFields}
D.~T. Hristopulos, Gaussian Random Fields, Springer Netherlands, Dordrecht,
  2020, pp. 245--307.

\bibitem{ChristoffelSymbols}
G.~Arfken, Mathematical Methods for Physicists, 3rd Edition, Academic Press,
  1985.

\bibitem{RK4}
J.~C. Butcher, Numerical Methods for Ordinary Differential Equations, John
  Wiley \& Sons, 2008.

\bibitem{MCMC2021}
G.~L. Jones, Q.~Qin, Markov chain monte carlo in practice, Annual Review of
  Statistics and Its Application 9~(1) (2021) null.

\bibitem{GMRFBook}
H.~Rue, L.~Held, Gaussian Markov Random Fields: Theory and Applications,
  Chapman \& Hall/CRC, 2005.

\bibitem{Hammersley}
J.~M. Hammersley, P.~Clifford,
  \href{www.statslab.cam.ac.uk/~grg/books/hammfest/hamm-cliff.pdf}{Markov field
  on finite graphs and lattices (preprint)} (1971).
\newline\urlprefix\url{www.statslab.cam.ac.uk/~grg/books/hammfest/hamm-cliff.pdf}

\bibitem{Besag}
J.~Besag, Spatial interaction and the statistical analysis of lattice systems,
  Journal of the Royal Statistical Society. Series B (Methodological) 36~(2)
  (1974) 192--236.

\bibitem{ElementaryDG}
B.~O'Neill, Elementary Differential Geometry, 2nd Edition, Elsevier, 2006.

\bibitem{Bar}
C.~B\"ar, Elementary Differential Geometry, Cambridge University Press, 2010.

\bibitem{Manfredo}
M.~P. do~Carmo, Differential Geometry of Curves and Surfaces, 2nd Edition,
  Dover Publications Inc., 2017.

\bibitem{Pressley}
A.~Pressley, Elementary Differential Geometry, 2nd Edition, Springer, 2012.

\bibitem{Woodward}
L.~Woodward, J.~Bolton, A First Course in Differential Geometry: Surfaces in
  Euclidean Space, Cambridge University Press, 2019.

\bibitem{Amari}
S.~Amari, H.~Nagaoka, Methods of Information Geometry, American Mathematical
  Society, 2000.

\bibitem{Nielsen}
F.~Nielsen, \href{https://www.mdpi.com/1099-4300/22/10/1100}{An elementary
  introduction to information geometry}, Entropy 22~(10) (2020) 1100.
\newblock \href {http://dx.doi.org/10.3390/e22101100}
  {\path{doi:10.3390/e22101100}}.
\newline\urlprefix\url{https://www.mdpi.com/1099-4300/22/10/1100}

\bibitem{isserlis1918}
L.~Isserlis, On a formula for the product-moment coefficient of any order of a
  normal frequency distribution in any number of variables, Biometrika 12
  (1918) 134--139.

\bibitem{Riemann}
J.~M. Lee, Riemannian manifolds: an introduction to curvature, Springer, 1997.

\end{thebibliography}

\end{document}